\documentclass[fleqn,usenatbib]{mnras}

\usepackage{newtxtext,newtxmath}
\usepackage[T1]{fontenc}
\usepackage{amsmath}
\usepackage{graphicx}
\usepackage{natbib}
\usepackage{CJKutf8}
\usepackage{xcolor}
\usepackage{times}
\usepackage{multirow}
\usepackage[normalem]{ulem}
\usepackage{url}

\title[Future Continuum Cosmology Clustering Surveys]{Future Radio Continuum Cosmology Clustering Surveys}
\author[Asorey and Parkinson]{Jacobo Asorey$^{1,2}$ and David Parkinson$^{2,3}$\thanks{E-mail:davidparkinson@kasi.re.kr}\\
$^1$ Centro de Investigaciones Energeticas, Medioambientales y Tecnologicas (CIEMAT), Av. Complutense, 40, 28040 Madrid, Spain\\ 
$^2$ Korea Astronomy and Space Science Institute, Yuseong-gu, Daedeok-daero 776, Daejeon 34055, Korea\\
$^3$University  of  Science  and  Technology,  Daejeon  34113,  Korea}
\date{Accepted XXX. Received YYY; in original form ZZZ}

\pubyear{2020}
\begin{document}
\label{firstpage}
\pagerange{\pageref{firstpage}--\pageref{lastpage}}
\maketitle

\maketitle

\begin{abstract}
The use of continuum emission radio galaxies as cosmological tracers of the large-scale structure will soon move into a new phase. Upcoming surveys from the Australian Square Kilometre Array Pathfinder (ASKAP), MeerKAT, and the Square Kilometre Array project (SKA) will survey the entire available sky down to an $\sim 100\mu$Jy flux limit, increasing the number of detected extra-galactic radio sources by several orders of magnitude.  External data and machine learning algorithms will also enable some low resolution radial selection (photometric redshift binning) of the sample, increasing the cosmological utility of the sample observed. In this paper, we discuss the flux limit required to detect enough galaxies to decrease the shot noise term in the error to be 10\% of the total. We show how future surveys of this type will be limited by available technology. The confusion generated by the intrinsic sizes of galaxies may have the consequence that surveys of this type eventually reach a hard flux limit of $\sim 100$nJy, as is predicted by the current modelling of AGN sizes by simulations such as the Tiered Radio Extragalactic Continuum Simulation (T-RECS). Finally, when considering the multi-tracer approach, where galaxies are split by type to measure some bias ratio, we find that there are not enough AGN present to achieve a reasonable level of shot noise for this kind of measurement.
\end{abstract}
\begin{keywords}
radio continuum: galaxies -- large-scale structure of Universe 
\end{keywords}

\section{Introduction}

The distribution of matter in the Universe, originally seeded as tiny fluctuations during the initial expansion, provide an important cosmological probe of both the physics of the early time, and the subsequent expansion and evolution. The late-time\footnote{Here `late-time' commonly refers to any survey or probe sampling the Universe after the cosmic microwave background was emitted at recombination. In this regard even an AGN survey extending up to redshift $z\sim5$ can be considered a late-time cosmology experiment.} matter distribution is often measured through the statistics of luminous matter, and this is most commonly in the form of galaxies.

The use of clustering statistics of continuum radio sources as a cosmological probe \citep{10.1046/j.1365-8711.2002.05163.x,SKA2015Jarvis,2016A&A...591A.135C} has lagged somewhat behind that of optical galaxies. There are two reasons for this: the relative difficulty of obtaining a equivalently large sample (given the available technology), and the comparative lack of spectral features at those wavelengths, making  localisation in the radial direction much harder. To close this gap, the next generation of radio observatories will need to have a high sensitivity and capable of both precise angular localisation and fast coverage of a wide-area. Fortunately, with the development of the SKA pathfinders such as ASKAP (the Australian Square Kilometre Array Pathfinder, \citet{askap2021}), MWA (Murchison Widefield Array, \citep{mwa2013}) and MeerKAT, this will be possible. The next few years should see a rapid growth in the number of galactic radio sources useful for cosmology as wide-area surveys reach an rms noise of $\sim 10 \mu\mathrm{Jy}$/beam \citep{2011PASA...28..215N,GLEAM_catalogue,MeerKLASS}.

However, while this door of opportunity may be opening now, it may not be possible to keep it open into the distant future. In attempting to reach sub $\mu$Jy flux limits for the next generation of instruments, all-sky radio continuum surveys beyond SKA Phase I may find the accessible area of the sky to be reduced. In the SKA technical memo by \cite{Condon2009Memo1S},  the author discussed the  effect of \textit{confusion noise} and \textit{natural confusion} providing a hard bound on the achievable flux limit of the survey, given the radio telescope being used. Even if this very faint limit can be reached for small, pinpoint surveys, extending these to the whole sky will be limited by the necessity of masking bright radio sources, and the presence of diffuse galactic  foreground emission increasing the difficulty of detecting faint sources. 

In \cite{2000ExA....10..419H} the authors made a detailed analysis of what technology would be required to detect radio continuum galaxies in the sub-$\mu$ Jy regime. In this paper, we build on those ideas and subsequent work to explore the optimal manner of sampling both wide areas and reaching deep flux limits that will be required for the cosmology use of radio-continuum surveys. Much of this work is based on the simulation of the radio sky from the Tiered Radio Extragalactic Continuum Simulation (T-RECS) \citep{2019MNRAS.482....2B}. Though this is the current `state of the art' simulation, it is still based on observational data. This means that the extrapolations that we make to as yet unexplored regions of parameter space (e.g. flux densities $\ll 1\mu$Jy) may not be reflected in future observations.

In section \ref{sec:clustering} we show how the flux limit for radio continuum surveys influences the measurement of the angular clustering statistics, and  how the separation of the sample into redshift bins and by population can reduce the precision of these clustering measurements. In section \ref{sec:sky_fraction} we show how achieving the maximal area that minimises sample variance can be technically challenging due to the sources of noise and confusion present in the sky. In section \ref{sec:design_options} we discuss the designs for current and future wide-area continuum surveys, and the instrumentation and data processing requirements to reach the proposed cosmology limits.

\section{Clustering surveys}
\label{sec:clustering}

The design of a future radio continuum survey will depend on previous surveys and the science goals that we have prioritised. In the area of cosmology the science topics of the mysterious acceleration of the Universe (through some possible dark energy or modified gravity), and the initial conditions of the early Universe (generated by cosmological inflation or some other mechanism) will both return optimal results from all-sky extra-galactic surveys. Previous work has identified that testing inflation through non-Gaussianity will require an all-sky survey, with a number density high enough for shot-noise effects to be sub-dominant, that can be sub-divided into at least five redshift bins \citep{10.1111/j.1365-2966.2012.20634.x,Raccanelli_2015,2019JCAP...02..030B}.   We set this as the standard required to measure the non-Gaussian component on the initial curvature fluctuation to the same or better accuracy than Planck.

In this section we discuss how the measured clustering power spectrum and its errors change as the survey achieves deeper flux limits, and the sample is sub-divided into redshift bins.

\subsection{Clustering statistics}
\label{sec:clusteringstats}

The two-point distribution of radio galaxy positions in angular space can be decomposed into a series of spherical harmonics. For a given direction on the sky $\vec{\theta}$, if the continuous density field in that direction $\sigma(\vec{\theta})$ is Gaussian and randomly distributed, then it can be decomposed into its multiple moment using spherical harmonics $Y_{\ell m}$, such that the amplitudes $a_{\ell m}$ are given by
\begin{equation}
a_{\ell m} = \int d\vec{\theta} Y^*_{\ell m} \sigma(\vec{\theta})\,.
\end{equation}
The amplitudes of the spherical harmonic contributions for each of the multipole moments $\ell$ and $m$ can be averaged in one of the two spherical directions (assuming istoropy) giving the  angular correlation power spectrum $C_{\ell}$, such that
\begin{equation}
\label{eqn:cl1}
  C_{\ell} = \langle a_{\ell m} a_{\ell m}^{*}\rangle\,.
\end{equation}
Assuming that the galaxies trace some underlying matter distribution, which can be represented as a power spectrum of fluctuations, then the prediction for the angular power spectrum from theory is given by
\begin{equation}
\label{eqn:cl2}
  C_{\ell}  = 4\pi \int \frac{dk}{k} \Delta^2(k)[ W_\ell^g(k)]^2\,,
\end{equation}
where $k$ is the wavenumber,  $\Delta^2(k)$ is the logarithmic matter power spectrum and $W^g_{\ell}(k)$ is the radio galaxy window function. This form assumes a single redshift bin with the galaxies distributed across this bin, giving the window function as \citep[e.g.][]{2008PhRvD..77l3520G,2008MNRAS.386.2161R},
\begin{equation}
\label{eqn:wincl}
  W_\ell(k) = \int \frac{dN(\chi)}{d\chi} b(z)D(z)j_\ell[k\chi]d\chi\,.
\end{equation}
Here $dN/d\chi$ is distribution of sources per steradian with comoving radial distance $\chi(z)$ within the redshift bin (brighter than some survey magnitude or flux limit), $b(\chi)$ is the bias factor relating tracer overdensity to matter overdensity, $D(\chi)$ is the growth factor of density perturbations, and $j_\ell(x)$ is the spherical Bessel function.

The power spectrum of density fluctuations $\Delta(k)$ and the cosmological distances $\chi(z)$ are sensitive to the values of the cosmological parameters and the cosmological model. With good knowledge of the bias and number distribution of radio galaxies, accurate measurements of the radio galaxy angular power spectrum can be used to constrain the cosmological parameters, and test the standard concordance cosmology $\Lambda$CDM. But in order to do so we will need very accurate measurement of this power spectrum.

We assume that the field describing the inhomogeneities is gaussian-distributed in the amplitudes, and so we can describe the covariance of the observed $C_{\ell}$ as \citep{Asorey2012}:
\begin{equation}
\mathrm{Cov}(C_\ell)= \frac{2(C_\ell+1/\bar{n})^2}{N(\ell)} \,,   
\end{equation}
where $N(\ell)=(2\ell+1)f_{\mathrm{sky}}$ is the number of modes sampled for a given $\ell$, for $f_{\rm sky}$ as the fraction of the sky being surveyed, and $\bar{n}$ is the angular number density of sources (in units of number count per steradian) given by $\bar{n}=N_{\mathrm{gal},\mathrm{bin}}/\Delta\Omega$. 

Therefore, the variance on an angular power spectrum measurement, in the completely Gaussian case, is given by 
\begin{equation}
    \label{eqn:cl_error}
    \sigma^2_{C_{\ell}} = \frac{2}{f_{\rm sky}(2\ell+1)}\left[C_{\ell}+\frac{1}{\bar{n}}\right]^2\,.
\end{equation}
The two sources of error can therefore be broken down into \textit{sample variance}, which is limited by the number of independent modes on the sky that the power can be measured in, and \textit{shot noise}, which is limited by the number density of sources. Finally, the signal in the angular power spectrum can also be increased by measuring tracers with a greater \textit{galaxy bias}, which will therefore have a larger angular clustering amplitude.

For a target sample of radio continuum galaxies, detected through their synchrotron emission, the number of sources will increase as the flux-limit decreases, and the redshift distribution will change. As future radio surveys reach a greater depth (smaller flux limit), equation \ref{eqn:cl_error} implies that the number density of sources will reach a sufficient number density such that the shot noise will be sub-dominant to the sample variance error. Using a model of the total number and redshift distribution of sources, we can compute the corresponding flux limit that gives a fixed fraction of the shot noise contribution to the total errors. 

\begin{figure*}
    \centering
    \includegraphics[width=0.90\textwidth]{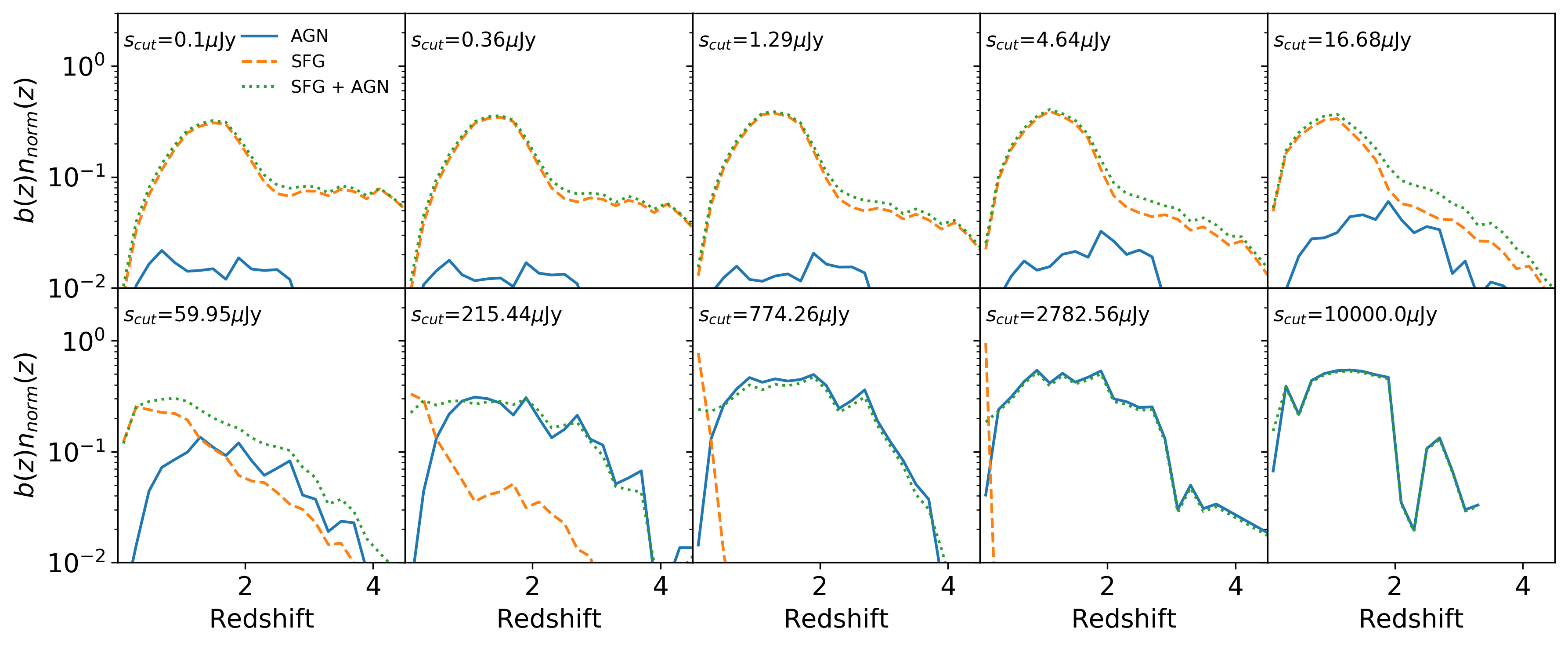}
    \caption{The evolution of the product of the bias and the normalised redshift distribution of total radio continuum populations, AGNs and SFGs for different survey flux cuts. Here the number distribution is drawn from the T-RECS simulation \citep{2019MNRAS.482....2B}, and the bias is estimated using the Colossus code \citep{2018ApJS..239...35D} with the relationship between halo mass and type the same as adopted by the $S³$ simulations \citep{wilman2008}. This quantity is averaged over when computing the theoretical value of the angular power spectrum, $C_\ell$, and the change in amplitude and localisation of $n(z)b(z)$ relates directly to the amplitude and shape of the $C_\ell$ spectrum. }
    \label{fig:redshift_number_bias}
\end{figure*}

The predicted normalised number distribution of galaxies, multiplied by the linear bias of that tracer, is the kernel that we integrate over to find the window function in Eqn.~\ref{eqn:wincl}, and is shown in Fig.~\ref{fig:redshift_number_bias}. In this figure we see that the overall amplitude changes as the number of less-biased SFGs come to dominate over the more-highly biased AGN. But the localisation changes as well, as we detect more high redshift SFGs at lower fluxes.

To extract cosmological-relevant data from the angular power spectrum we need to be able to both accurately measure and accurately model it, which gives a limit to range of multipoles we can utilise, as modelling the non-linear power-spectrum accurately becomes more difficult as $k$ increases. Following the approach in \cite{2019JCAP...02..030B}, we define a minimum and maximum multipole number. For the largest scales, $\ell_{\rm min}$ is limited by the fraction of sky surveyed, $\ell_{\rm min} = \pi/(2f_{\rm sky})$.  We assume a 75\% total sky coverage, though address the question of sky fraction as a function of flux limit in section \ref{sec:sky_fraction}. For the small-scale limit we need to consider the distance, $\chi(z)$ to the particular redshift bin in consideration. We define the relevant scale $r_\star$ as
\begin{equation}
  r_\star = \int{d\chi\frac{dN(\chi)}{d\chi}} 
  \label{eqn:r_max}
\end{equation}
and then we define $\ell_{\rm max}=k_{\mathrm{max}}r_\star$, where $k_{\mathrm{max}}=0.1$ in order to consider only the linear regimes.

In order to realistically simulate the predicted $C_\ell$ and  shot noise, we need to model the galaxy bias $b(z)$ and the redshift distribution $N(z)$ of the samples we will observe, for a given flux cut. We use a combination of information from two sets of simulations. We define the redshift distributions of a given population with the T-RECS simulation \citep{2019MNRAS.482....2B}. However, for the galaxy bias we use  the Colossus \citep{2018ApJS..239...35D} suite to estimate the bias from the halo model, using the Tinker model \citep{2008ApJ...688..709T}. The values for the halo masses assigned to each type of radio galaxy taken by the $S³$ simulations \citep{wilman2008}, which follows the approach of \cite{10.1111/j.1365-2966.2012.20634.x} and \cite{2019JCAP...02..030B}, associating the names in that work with the names given by T-RECS \citep{2019MNRAS.482....2B}. In Tab.~\ref{tab:radiogalclass} we show the different names given to different classifications of radio galaxies for different works. 

\begin{table*}
\caption{\label{tab:radiogalclass} A comparison given to the names of different types of radio galaxies, based on the way that they are classified.}
\label{table:tab_types_gals}
\begin{tabular}{ l  l  l }
\hline \hline
SKADS name  & Spectral classification  & T-RECS name \\
\citep{wilman2008} & \citep{Hale2018}  & \citep{2019MNRAS.482....2B}  \\
\hline
Star-forming galaxy (SFG) & \multirow{ 2}{*}{SFG} & {SFG} \\
Starburst (SB) & & SB \\
RQQ (Radio Quiet Quasar) &  \multirow{ 2}{*}{Moderate to low Luminosity AGN (MLAGN)} &  BL Lac\\
FRI (Fanaroff-Riley type I) &   & Flat-spectrum radio quasar (FSRQ) \\
FRII (Fanaroff-Riley type II) &High to moderate  Luminosity  AGN (HLAGN) &   Steep-spectrum AGN (SS-AGN) \\
\hline
\end{tabular}

\end{table*}

As the relevant quantity is the combination of the bias $b(z)n(z)$, we need to compute a combined bias $b(z)$ for the ensemble. We combine the different populations in the following way:
\begin{equation}
    b(z) = \frac{\sum_\alpha^{\mathrm{type}}{b_\alpha(z)N_\alpha(z)}}{N_{all}(z)}\,,
\end{equation}
where $\alpha = \{AGN_{\mathrm{RQQ}}$,~$AGN_{FRI}$,~$AGN_{FRII}$,~$SFG_{\mathrm{SFG}}$,~$SFG_{\mathrm{SB}}\}$ following the types used in the $S^3$ simulation. We derive this form of the total bias of the ensemble in appendix \ref{appendix:bias_total}. As seen in Eq. \ref{eqn:wincl}, the combination of the bias and the normalized redshift distribution $b(z)n_{\mathrm{norm}}(z)$ is the main component of the angular power spectrum window function. We show in Figure \ref{fig:redshift_number_bias} the kernel of this window function for different flux cuts. For bright flux limits, where the redshift distribution is dominated by AGN, there is a inherent noisiness to the redshift distribution of this species, caused by the small number of galaxies and the finite size of the simulation. However,  this distribution is firstly transformed through  Eqn.~\ref{eqn:wincl} to get the window function, and the square of the window function is convolved with the three-dimensional matter power spectrum to give the  angular power spectrum through Eqn.~\ref{eqn:cl2}, and both operations will smooth over any initial oscillations in  $n(z)$. We therefore expect this noisiness to have almost no effect on the predicted power spectrum at bright flux limits, and no consequence for our conclusions.

\begin{figure}
    \centering
    \includegraphics[width=0.45\textwidth]{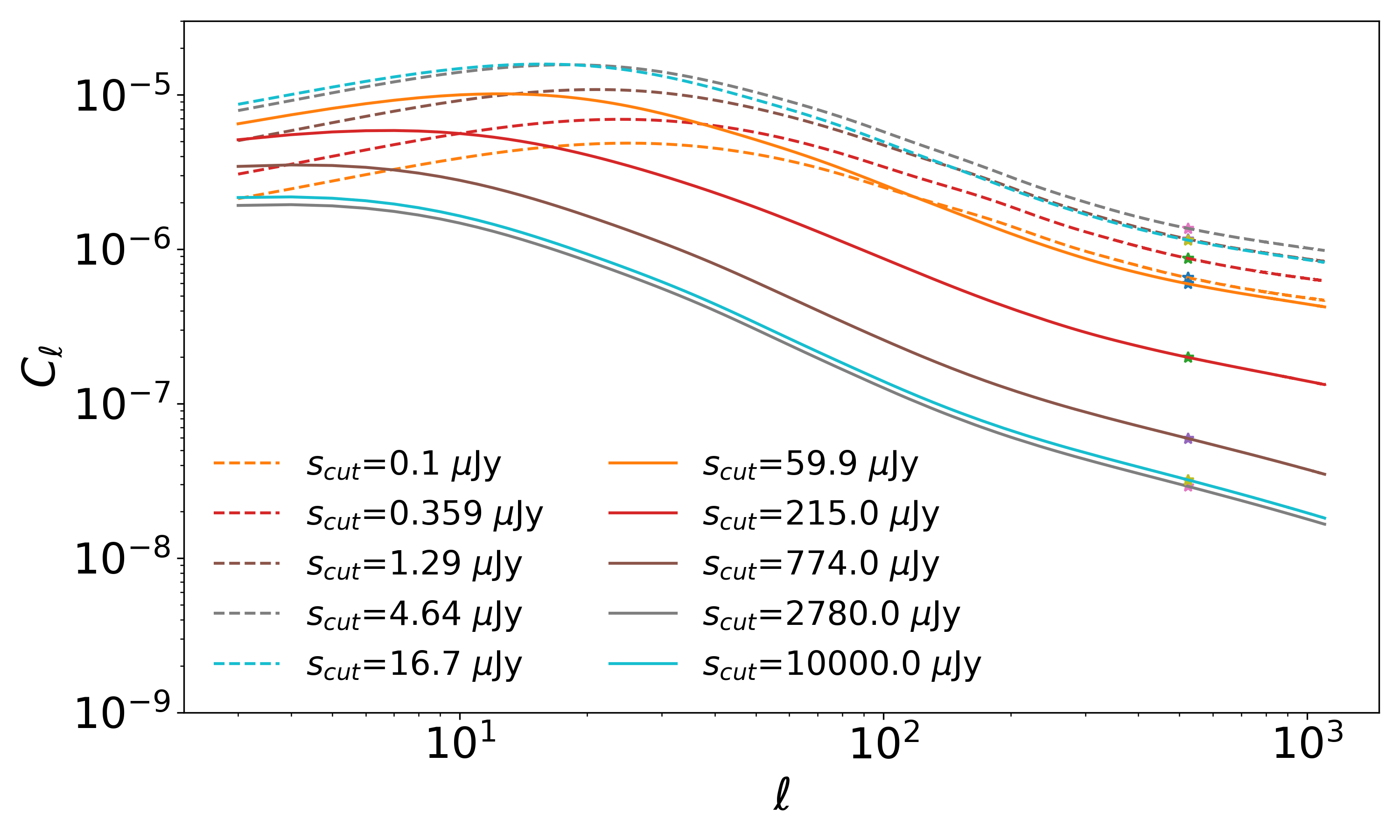}\\
    \includegraphics[width=0.45\textwidth]{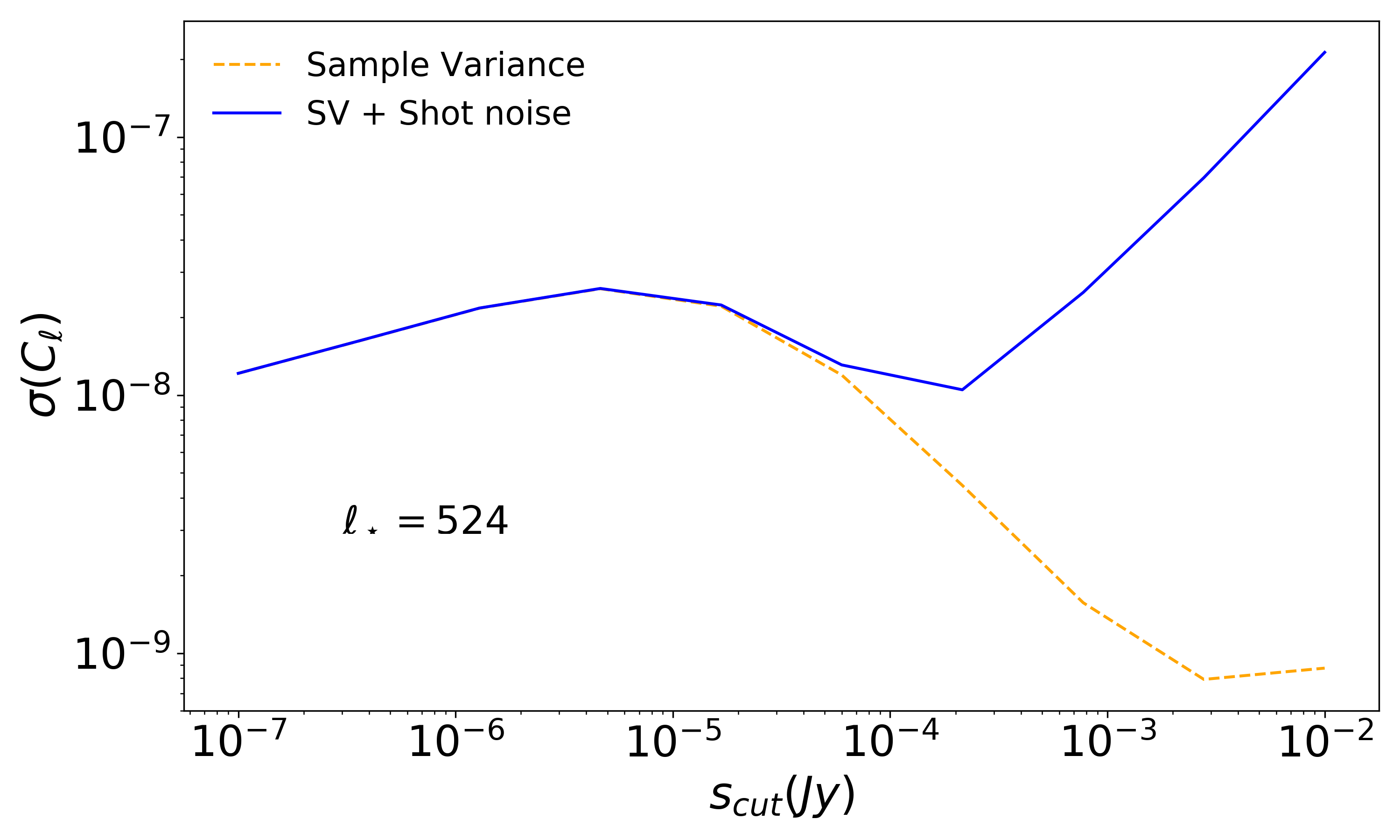}
    \caption{Predicted angular power spectra $C_{\ell}$ \textit{(top)} and error on the angular power spectrum \textit{(bottom)} of the clustering of radio continuum galaxies of a single redshift bin, for different values of the flux limit $s_{\mathrm{cut}}$. We see the increase and decrease in overall power amplitude as the flux limit changes, as well as a relatively small change in shape, owing to the number distribution and bias evolution of the tracers. For the error, the sample variance contribution in orange, and the combined sample variance and shot noise effect combined in blue. The sample variance error increases on the right hand side as the amplitude also increases with decreasing $s_{\mathrm{cut}}$, \label{fig:cl_flux_limit} }
\end{figure}

In figure \ref{fig:cl_flux_limit} we show how the angular power spectra changes with flux limit $S_{\rm cut}$. We find that as the flux limit decreases and more sources are detected, the power increases until a maximum at around $S_{\rm cut} \sim 10^{-5} $Jy, owing to the increased radial localisation of the tracers, as the number of low-redshift AGN and SFG rises, as well as a higher overall bias. For flux limits smaller than this, the amplitude decreases again slightly, as the number of higher redshift tracers increases, and the power becomes more radially smeared.

In figure \ref{fig:cl_flux_limit} we also show how the error on the angular power spectrum 
changes with flux limit, for the sample variance alone, and the sample variance and shot noise combined. We find that the shot noise contribution to the error becomes sub-dominant to the sample variance at a flux limit of roughly $30\mu$Jy, and the two lines converge. Note that since the shot noise component is inversely proportional  to number density, rather than raw number, this change is independent of the total area surveyed. If all the galaxies that can be detected are used in a single `redshift bin', then integrating down to a flux limit fainter than $30\mu$Jy will increase the number of galaxies, but will not decrease the fractional  error on the measured $C_{\ell}$ power spectrum.

\subsection{Redshift binning}
\label{sec:redshiftbinning}

\begin{figure}
    \centering
    \includegraphics[width=0.45\textwidth]{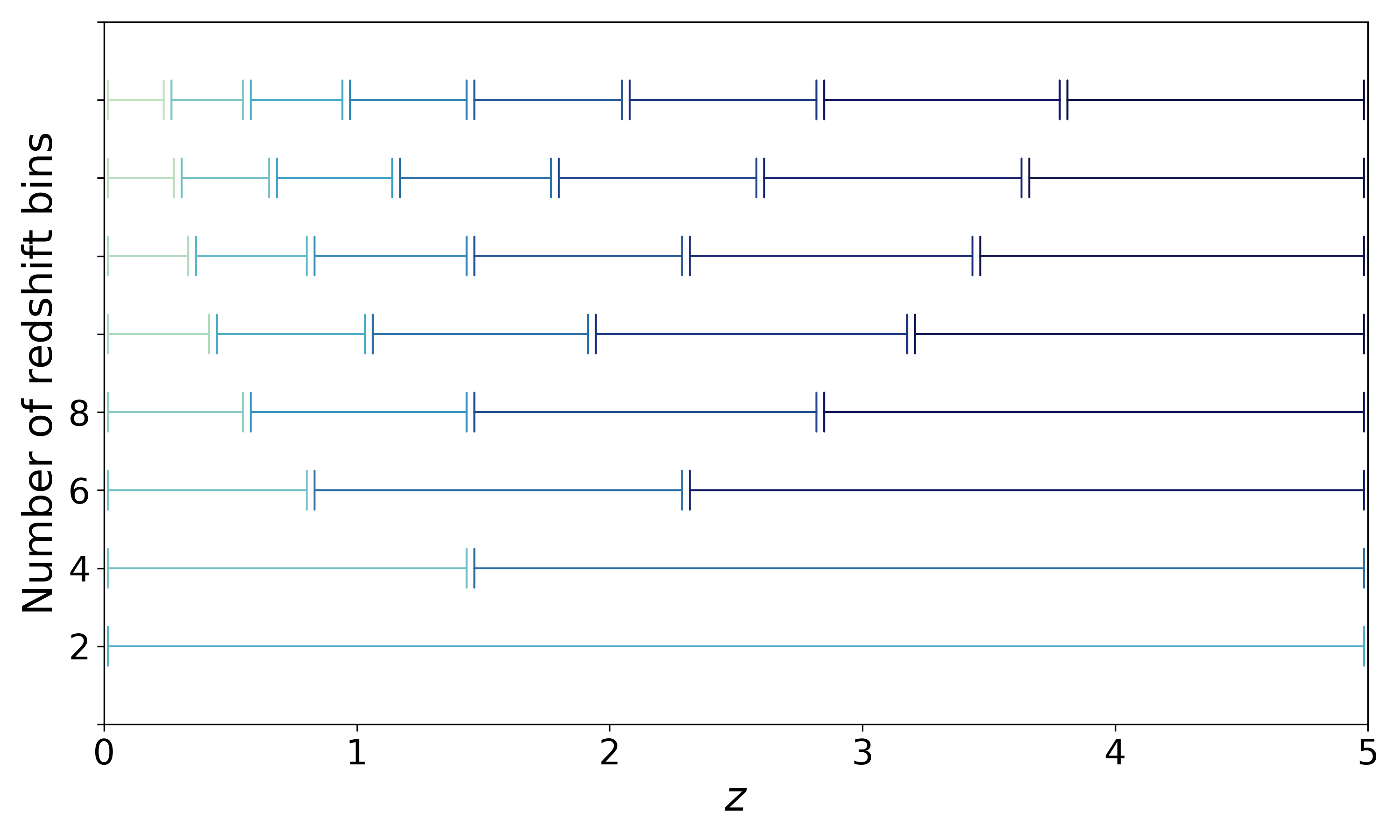}
    \caption{The ranges in redshift of the different redshift binning configuration considered in this paper. We assumed that there would be a greater amount of multi-spectrum, morphological, and clustering information available at lower redshifts than higher, which would allow more precise sub-divisions in redshift. In this paper we assume this would manifest as for equal binning in $\log(1+z)$, as shown here. Throughout this paper we assume top-hat redshift bins.}
    \label{fig:redshift_binning}
\end{figure}

Sub-dividing the sample of galaxies by redshift will allow the angular power spectrum to be measured in more than one redshift bin. However, as continuum sources are normally detected at the $\approx 1$GHz frequency through their synchrotron emission, the featureless power-law nature of their flux spectrum makes for a greater challenge than for optical photometric redshift binning. As such, future surveys may need to make use of data from other surveys, to either cross-identify with an optical/NIR source with redshift information, or make use of statistical clustering redshifts \citep{10.1093/mnras/stx691}, again using known redshifts of sources over the same area of sky. In either case though, much more information is needed to sub-divide the sample by redshift will be available at low redshift than high. In this paper, rather than assuming a particular set of redshift bins, we assume that the sub-division will be roughly equal in $\log(1+z)$, with a maximum of $z=5$. In figure \ref{fig:redshift_binning} we show the redshift ranges spanned by the different bins for each redshift binning case considered in this paper.

As the galaxies being used in the sample are localised into smaller bins, the minimum scale that can accurately be modelled also changes, and will be different for each bin. We compute the maximum multiple number $\ell_*$ using Eqn.~\ref{eqn:r_max} for each bin. We show the values of $l_*$ for the extreme case of eight redshift bins in Tab.~\ref{tab:lmax_bins}.

\begin{table}
\caption{\label{tab:lmax_bins} Scales used in the analysis for the configuration with 8 redshift bins that correspond to a comoving scale of $k=0.1$ Mpc$^{-1}$. }
\centering
\begin{tabular}{ l  l  l }
\hline \hline
Redshift bin  & Mean distance  & $\ell_\star$ \\
\hline
0-0.25 & 504 &50 \\
0.25-0.57 &1546 &155 \\
0.57-0.96 &2649   &265  \\
0.96-1.45 &3756   &376  \\
1.45-2.06 &4822 &482   \\
2.06-2.83 &5815 &581   \\
2.83-3.8 &6725 & 673  \\
3.8-5 &7548 & 755  \\
\hline
\end{tabular}

\end{table}

\begin{figure*}
    \centering
    \includegraphics[width=0.45\textwidth]{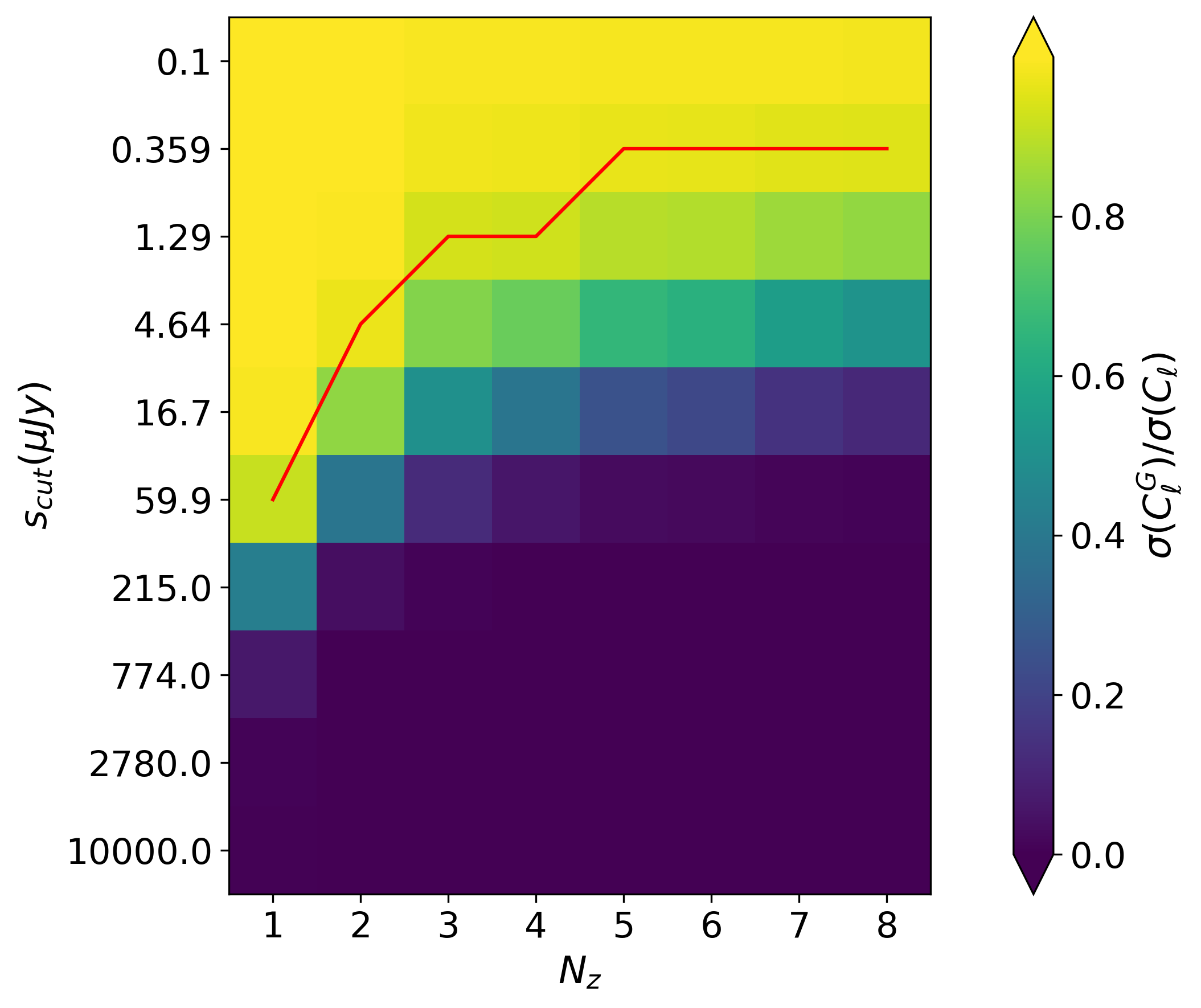}
    \includegraphics[width=0.45\textwidth]{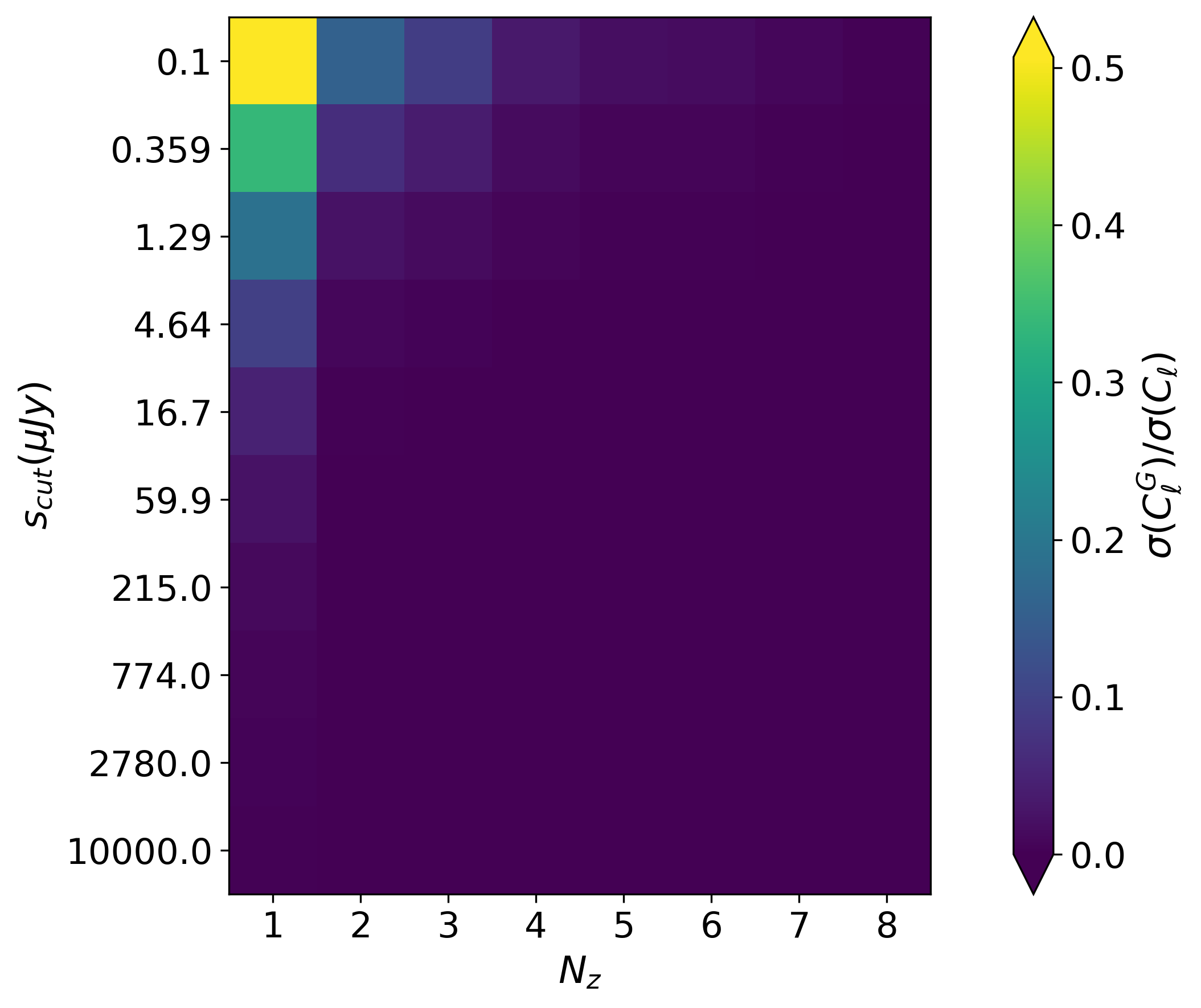}
    \caption{Fraction of the signal noise that is shot noise for different redshift configurations (described by the number of redshift bins $n_{\rm bins}$) and flux limits, for all galaxies (\textit{left}) and AGN only (\textit{right}). The fraction is shown by the colour bar, with dark blue being mostly shot noise in the error budget, and bright yellow being mostly sample variance. The red line shows the flux limit that is needed for each redshift case for the shot noise to be less than $10\%$ of the total signal noise. There is no red-line on the right panel of the figure, because in no cases do we achieve a less than $10\%$ shot-noise contribution to the error budget when using AGN only. This assumes an all-sky survey, $f_{\mathrm{sky}}=1$}
    \label{fig:flux_limit_snr}
\end{figure*}

When the sample of galaxies can be sub-divided, either by redshift or population, or both, the required number of galaxies increases. We assume a number of redshift bins, with the size of each scaled by the logarithm of $1+z$ and spaced to cover the range $0<z<5$, as shown in figure \ref{fig:redshift_binning}.  In the left panel of figure \ref{fig:flux_limit_snr} we show how the flux limit for the total population required to achieve 10\% shot noise increases with number of bins. We find that there is a `plateau' between $n_{\rm bins}=5$ and $n_{\rm bins}=8$ for a flux limit of $0.5\mu$Jy, such that subdividing the sample into more bins does not significantly impact on the shot noise contribution.

\subsection{Multi-tracer sampling}
One way to remove or reduce sample variance consists of using two differently biased tracers of the same matter field \citep{2009JCAP...10..007M,2009PhRvL.102b1302S,2014MNRAS.445.2825A}, and measuring the ratio of the density fluctuation of the two different samples, over the same volume. If the two tracers are following the same underlying matter field (i.e. there is no stochasticity), it follows that this ratio will be independent of the underlying fluctuation. Then the error on the power spectrum measurement of this ratio will be independent of the modes of the underlying fluctuation, effectively \textit{cancelling} the sample variance, allowing for better than cosmic variance uncertainty on large-scale physical effects that are not dependent on the underlying power spectra, such as the non-linear bias generated by primordial non-Gaussianity (e.g. \cite{2019JCAP...02..030B,10.1093/mnras/stz3581}).

For this technique to be effective, a large number density of tracers with different biases must be surveyed over the entire survey volume. In the radio continuum it is possible to separate the sample into AGN and star-forming galaxies, though doing so will reduce the number density, in comparison to the combined cohort. To achieve the same shot-noise accuracy as we defined for Sec.~\ref{sec:redshiftbinning}, it will be necessary to increase the numbers of these individual populations by pushing down to even smaller flux limits.

In the right panel of figure \ref{fig:flux_limit_snr}, we show the flux limit required in order to avoid shot noise saturation of the error for a given set for redshift bins, when only considering AGN galaxies. We only consider AGN here because their number density is much lower than that of SFGs below 10mJy, as shown in Fig.~\ref{fig:redshift_number_bias}, and so this population will be the limiting factor. It is shown that we can only achieve the same shot noise error by reaching a flux limit of  $s_{\mathrm{cut}}=0.1\mu Jy$, and in this case only for a single continuous redshift bin. This is similar to the result found in \citet{2019JCAP...02..030B}, where the best constraint on the non-Gaussianity parameter $f_{\mathrm{NL}}$ from the EMU survey would come from the analysis with an SFG \& AGN multitracer approach but only a single redshift bin. Here we state more strongly that no future continuum surveys will be able to sub-divide in both redshift and tracer and see any improvement in cosmological constraints, even those achieving a fainter flux limit than EMU.

\section{On-sky radio survey limits}
\label{sec:sky_fraction}

If the design goal of large-area angular clustering surveys is to minimise the shot-noise contribution to the error that comes from a low number density of tracers, it must also be to minimise the sample variance by maximising the sky area covered. However, even if there is enough observing time to integrate down to the required flux limit uniformly over the entire available sky, the resulting sample of tracers may not have the same flux limit and sky fraction as was planned. This is because there are on-sky limitations to the effectiveness of the survey that are generated by bright sources, faint diffuse foreground emission, and confusion between tracers being resolved in the same beam. In this section we discuss each of them in turn.

\subsection{Foreground maps}
The signal in a given beam would be the combination of the source temperature and the system (instrument) noise:
\begin{equation}
T_A = T_{\mathrm{sys}} + T_{\mathrm{source}}
\end{equation}
where the system temperature is a combination of the background, which is mostly given by foreground emission ($T_{\mathrm{sky}}$), the atmospheric emission ($T_{\mathrm{atm}}$), ground signal scattered from the feed and support structure ($T_{\mathrm{scat}}$), and electronics noise from the receiver ($T_{\mathrm{recv}}$), i.e.
\begin{equation}
    T_{\mathrm{sys}} = T_{\mathrm{sky}} + T_{\mathrm{atm}} + T_{\mathrm{scat}} + T_{\mathrm{recv}}\,.
\end{equation}

The expected flux limit of a radio continuum survey is derived from the root-mean square noise on the flux intensity map. For a given integration on the sky of time $\tau$, the rms noise in the flux measurement ($S_{\mathrm{rms}}$) from the temperature of the system is given by the radiometer equation,
\begin{equation}
    \label{eqn:radiometer}
    S_{\mathrm{rms}} = \frac{2k_BT_{\mathrm{sys}}}{A_{\mathrm{eff}}N_{\rm dish}\sqrt{2B\tau}} \,,
\end{equation}
where $k_B$ is the Boltzmann constant, $B$ is the bandwidth, $A_{\mathrm{eff}}$ is the effective collecting area, and $N_{\rm dish}$ is the number of dishes. Note that this noise is assumed to be Gaussian and generated by the internal temperature fluctuations of the radio telescope and some homogeneous external radio field (background sky temperature). Sources are thus identified as being fluctuations of an amplitude significantly larger than this flux rms, and normally associated with a cut at (for example) 10-sigma. While random fluctuations  of this size can happen, the probability of such a fluctuation ($10^{-23}$)  in a sample of $10^{10}$ -- $10^{12}$ pixels is   small enough to avoid consideration.

The diffuse sky brightness can overwhelm the emission from a localised extragalactic source, reducing the signal to noise ratio of the detection. The closer to the galactic equator that observations are made, the brighter the sky will be, and longer integrations will be needed to achieve a uniform deepness. Reversing this argument, we can estimate the amount of sky that will be inaccessible for a given flux limit and integration time. 

\begin{figure}
    \centering
    \includegraphics[width=0.45\textwidth]{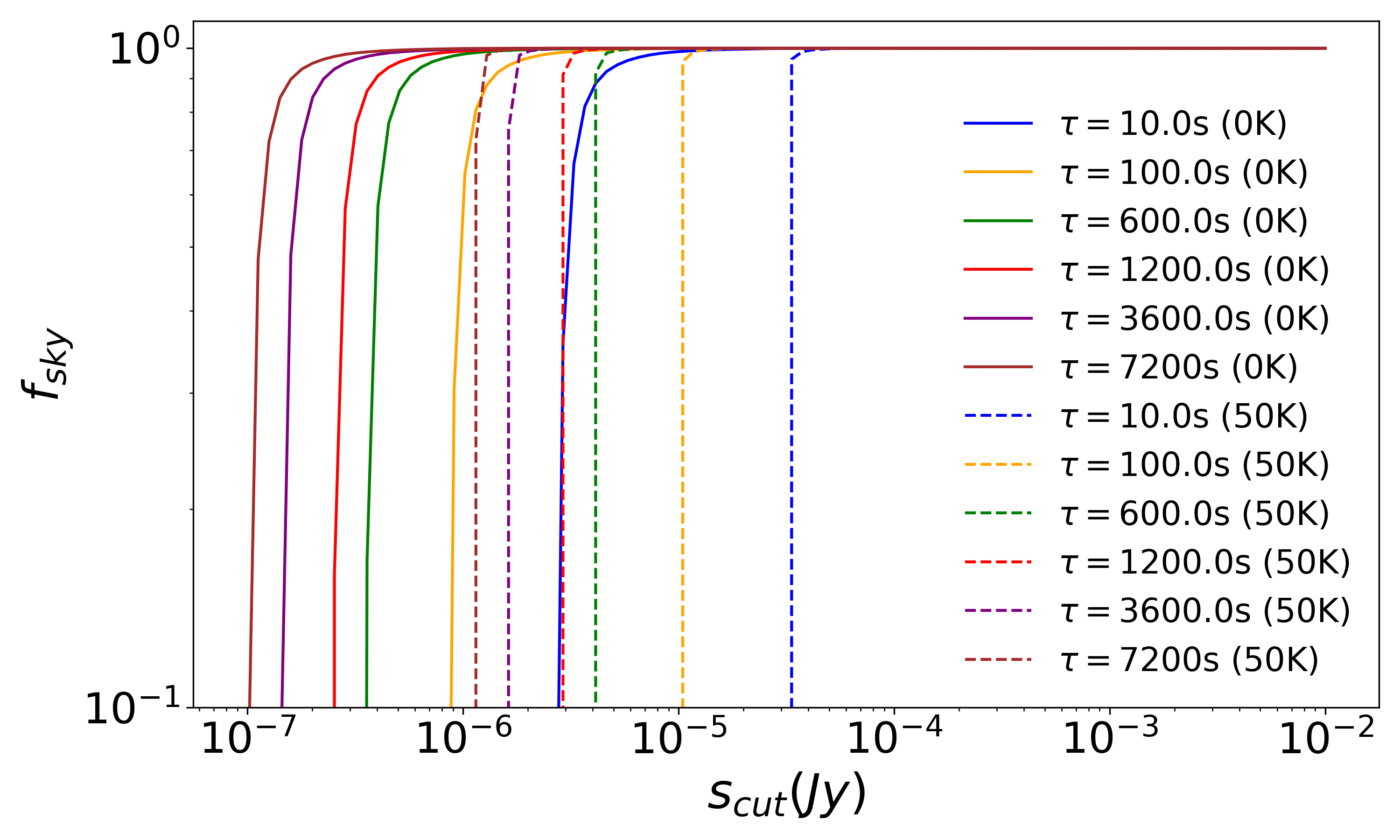}
    \caption{Fraction of the sky that will be inaccessible as a function of integration time ($\tau$) for different flux limits and different system temperatures, assuming a 10$\sigma$ detection threshold. We assume a background sky modelled by the Global Sky Model (GSM, \citet{2008MNRAS.388..247D,2017MNRAS.464.3486Z}), a frequency range of 800 MHz to 1.4 GHz, a brightness efficiency of $\eta_B=1.0$, and a beam size of 60 arcseconds FWHM.}
    \label{fig:fsky_fcut}
\end{figure}

We use the Global Sky Model \citep{2008MNRAS.388..247D,2017MNRAS.464.3486Z} to model the sky temperature across the sky, and the radiometer equation (Eqn.~\ref{eqn:radiometer}) to model the SNR for detection.
In figure \ref{fig:fsky_fcut} we show how the sky fraction varies with exposure time and flux limit, for two possible system temperatures. To reach the required depth across the sky, such that the number of galaxies gives $<10\%$ shot noise for at least two redshift bins, we would need integrations of at least two hours for a 50K detector. Greater depths can be reach with longer integrations or more sensitive equipment.

This argument is applied to the entire sky, not to any particular observing patch. There are, of course quieter regions of the sky that can be observed that have relatively low sky temperature (e.g. the Lockman Hole), and so can be used for deeper observations with shorter integration time. In cosmology, since the sample variance will decrease as the survey area increases, the aim will always be to maximise the areal coverage. For a fixed total observing time, there will therefore be a trade-off between longer integrations and a wider area.

\subsection{Confusion limits}

\begin{figure}
    \centering
    \includegraphics[width=0.45\textwidth]{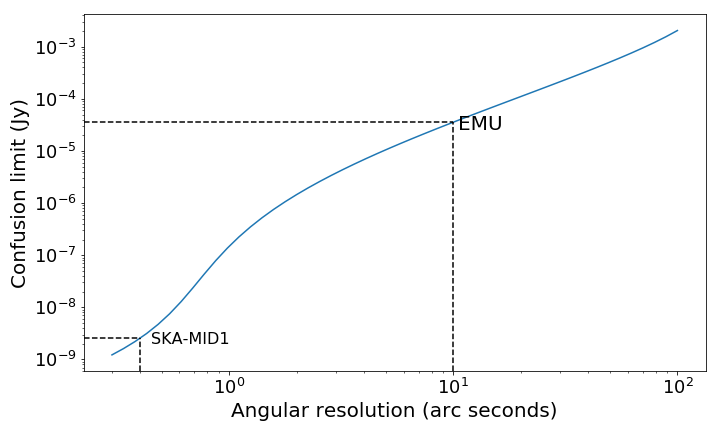}
    \caption{Flux limit of different surveys generated by the confusion of multiple detectable sources in the same beam, as a function of angular resolution. Here we assume a frequency of 1GHz. We see that the EMU survey, which has a confusion limit of $\sim30\mu$Jy ($5\sigma$) is operating close to the design limit given by the angular resolution, as it lacks the very long 150km baselines of SKA-MID1.}
    \label{fig:flux_confusion}
\end{figure}

Even if the integration time is increased indefinitely, there are natural limits to the depth at which radio continuum surveys can operate. The homogeneous radio sky approximation breaks down when considering the existence of faint extra-galactic radio sources, which generate `confusion noise'  \citep{2012ApJ...758...23C}. The number of sources above some flux limit is given by
\begin{equation}
N(>S_0) \int_{S_0}^\infty n(S) dS\,.
\end{equation}
The solid angle for a particular beam is 
\begin{equation}
    \Omega_{\rm beam}=\frac{\pi\theta^2}{4\ln(2)}\,,
\end{equation}
where $\theta$ is the beam FWHM. If the angular resolution of the telescope is too small, the number of sources per beam will be too large, and the images of these faint sources will overlap, causing the images generated to merge. So even if they are bright enough to be detected, the generated image will be too noisy for the sources to be distinguished. In this case the  receiver may be sensitive enough to detect the flux from the fainter sources, but the source detection is limited by the ability to separate them on the sky.  This limit is given by 
\begin{equation}
    \beta = [N(>S_0)\Omega_{\rm beam}]^{-1}\,,
\end{equation}
and is described in terms of some `number of sigma' cut, where the sources are distinct enough on the sky to be non-overlapping. So for $\beta=25$, this would a 5-sigma confusion limit for a flux cut of $S_0$, assuming a slope of 2 for the differential source counts.  We show how this translates into a flux limit for a given angular resolution, assuming the T-RECS simulated catalogue, in Fig.~\ref{fig:flux_confusion}.

At around 1GHz, we see that the EMU survey, which has a confusion limit of $\sim30\mu$Jy ($5\sigma$) is operating close to the design limit given by the angular resolution of ASKAP. SKA-MID1 does a lot better here, as it has access to some very long baselines of 150km. However, using these very long baselines and so generating  very high resolution images will increase the processing power required to run such a survey (though the discussion of processing power is beyond the current scope of the paper).

An increase in the number of long-baselines that is needed to achieve some high resolution imaging, and so lower the confusion limit, will also increase the computing power capacity needed to process the data. The ASKAP correlator is built to operate at 340 TFLOPS, and the raw data rate for ASKAP is approximately 100 Tbit/s \citep{askap2021}. However, the data rate would be expected to scale as $(B_{\mathrm{max}}/D)^4$, where $B_{\mathrm{max}}$ is the length of the maximum baseline, and $D$ is the diameter of the dish. If an observatory was constructed with similar size dishes to ASKAP but the 150km baselines planed for SKA Phase-I, the increase in data rate would be by a factor of $(150/6)^4 \sim 10^8 $, or approximately 100 exaFLOPS. This would be a more than significant upgrade in comparison to the current computational facilities.

This confusion noise is generated even in the `scale-free' case, where the clustering of these sources is not considered. There is also a `natural confusion' limit, where the inherent angular size of the sources themselves is large enough for them to overlap on the sky, and increased angular resolution will not save you from noise-limited images. In this case the limit is given by 
\begin{equation}
    \beta_{\rm n.c.} = [N(>S_0)\Omega_{\rm source}]^{-1}\,.
\end{equation}
If the source solid angle is given by $\Omega_s \sim \pi\theta^2_s/[4\ln(2)]$, where $\theta_s$ is the median angular source size, then we can compute the value of $\theta_s$ that corresponds to 5$\sigma$ natural confusion limit (again assuming a slope of 2 for the differential source counts). If the average source radius is above this size, then the sources will be overlapping and confused. 

\begin{figure}
    \centering
    \includegraphics[width=0.45\textwidth]{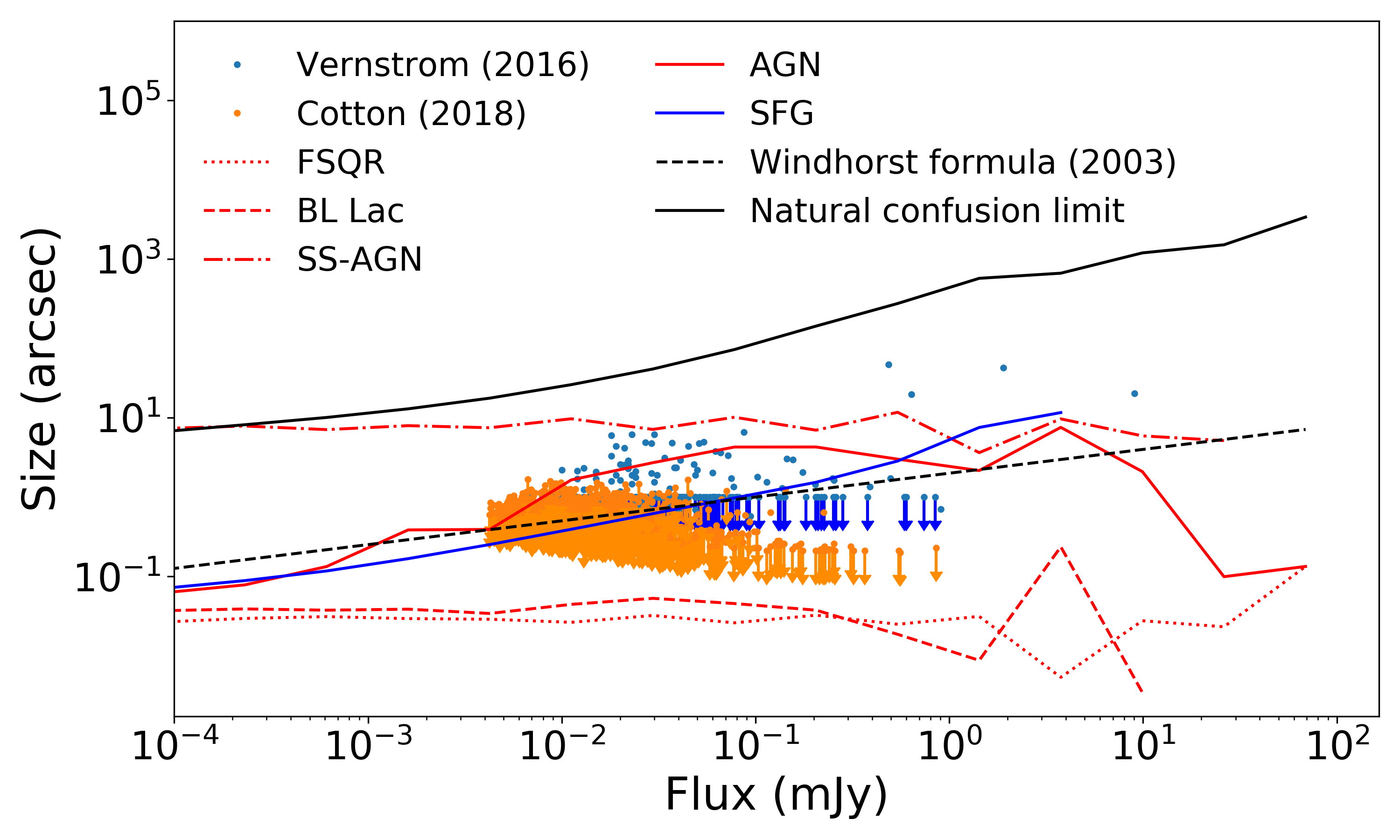}
    \includegraphics[width=0.45\textwidth]{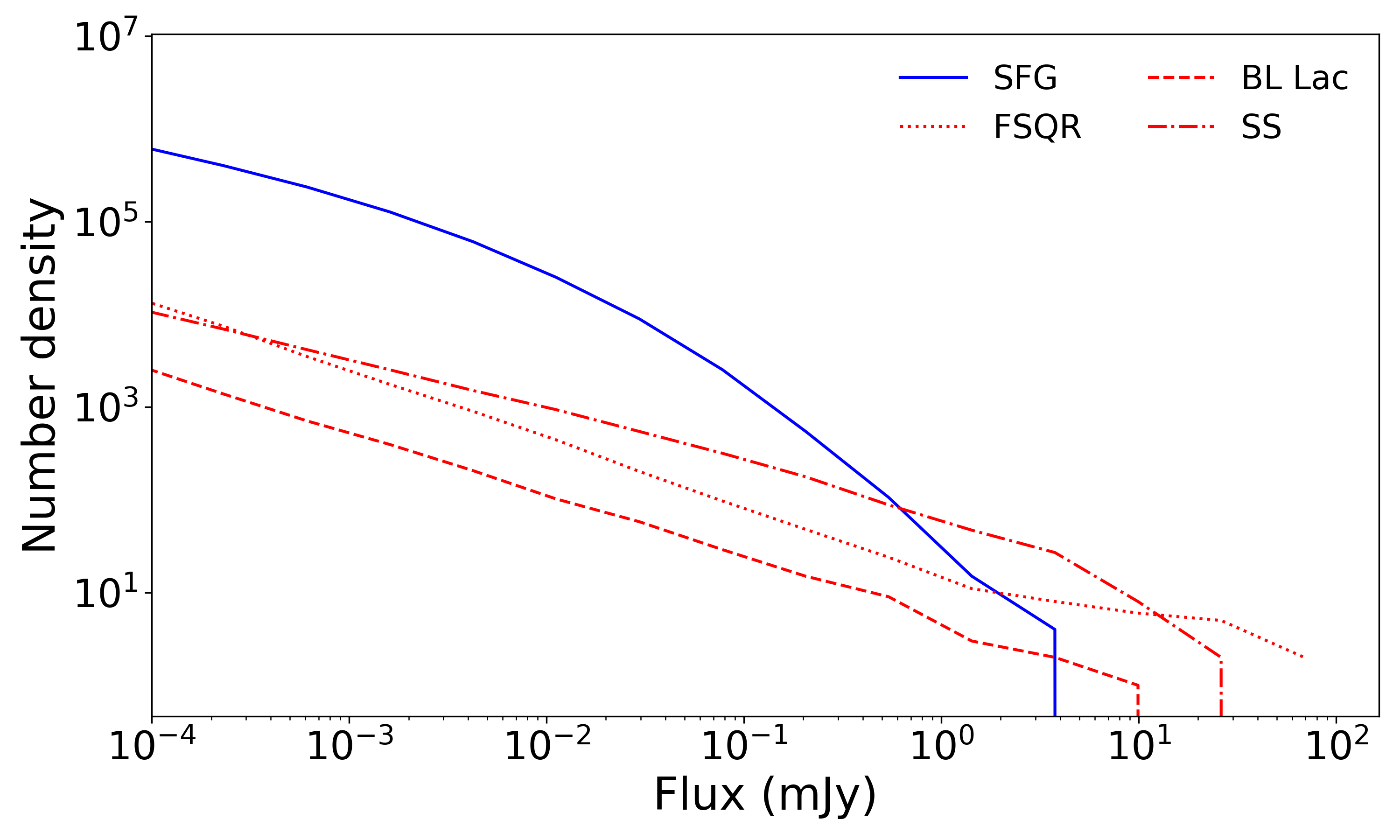}
   \caption{\label{fig:natural_confusion} The median size of radio continuum galaxies (\textit{top}), and the number of such objects (\textit{bottom}) as a function of flux limit, for AGN and SFG, as taken from the T-RECS deep simulated catalogue \protect\citep{2019MNRAS.482....2B}. We also add the sizes of galaxies (or the upper limits on their sizes) as measured by the \protect{\citet{Cotton2018}} and \protect{\citet{Vernstrom2016}} surveys at 3GHz.  We compare these sizes to the 5-$\sigma$ natural confusion limit and the best fit formula for the median source size taken from \protect\cite{2003NewAR..47..357W}. Here we assume a frequency of 1.4GHz. We find that at a flux limit of about 100 nJy the presence of a large number of `monster'  Steep-Spectrum AGN (SS-AGN) would place a limit on the possible depth of large-area surveys. }
\end{figure}

In Fig.~\ref{fig:natural_confusion} we show this source size limit as a function of flux limit, and compare with the source sizes present in the T-RECS catalogue \citep{2019MNRAS.482....2B}, for SFG and the different classifications of AGN present in the T-RECS simulated catalogue. We also compare these sizes to the best fit curve for the median sizes as a function of flux density taken from \cite{2003NewAR..47..357W}, which is given by
\begin{equation}
    \label{eqn:windhorst}
    \theta_s = 2\left(\frac{S}{1m\mathrm{Jy}}\right)^{0.3}\,,
\end{equation}
where $S$ is flux and $\theta_s$ is median source angular size, measured in arcseconds.  We also add the sizes of galaxies (or upper limit on sizes) as measured by the \protect{\citet{Cotton2018}} and \protect{\citet{Vernstrom2016}} surveys at 3GHz, showing that sizes have been measured down to roughly 10$\mu$Jy, but no deeper.

The Windhost relation relation is a reasonable fit to the SFG sizes from T-RECS for almost all flux densities. However, this is not the case for AGN.  The T-RECS simulation assumes a constant size-flux relation of the different AGN species. It therefore contains a number of `monster' AGN (the largest species, Steep-Spectrum AGN), which have large sizes but low flux densities. Though these would be subdominant in number, they would be large enough to confuse any surveys operating at the nJy level. If this population is present in reality, this would be a hard limit beyond which it would not be possible to increase the number density by further integration.

\subsection{Masking bright sources}

 For a given faint flux limit, there is a corresponding maximum flux limit beyond which bright sources will cause problems for reaching the required faint limit. The relationship between the bright and faint limits is given by the dynamic range $\mathrm{DR}$, such that
\begin{equation}
    \mathrm{DR} = 10\log_{10}\left(\frac{S_{\rm max}}{S_{\rm min}} \right)\,,
\end{equation}
where the dynamic range is given in decibels (dB).

Any sources in the field with a flux higher than this limit will generate imaging artefacts in the map, as the sidelobes of bright sources will be incompletely CLEANed \citep{1974A&AS...15..417H,2016A&A...591A.135C}, and so obscure nearby fainter sources. Therefore these will need to masked before the angular correlation function can be accurately measured. In the NVSS catalogue \citep{NVSS} sources closer than about 35' to a strong source of flux density $S$ and weaker than $10^{-3}S$ (given the local dynamic range assumed to be 30) were excluded from the analysis \citep{10.1046/j.1365-8711.2002.05163.x,2016A&A...591A.135C}. So as the dynamic range decreases the number of sources above the critical flux maximum $S_{\rm max}$ will increase, and,  because of the need to mask more of these bright sources, a greater area  of the sky will become unusable.

Different analysis have made different assumptions about the amount of area that needs to be cut. In \cite{2016A&A...591A.135C} they cut a circular region of radius 0.6$^o$ around the bright sources, which is an area of an area of roughly 1 sq. deg. But in \cite{2018JCAP...04..031B} this is widened to a 1$^o$ radius circle, which is an area of roughly 3 sq. deg. Even increasing the number or length of the baselines may not help reduce the size of the region to be removed, since, as shown in Fig.~\ref{fig:natural_confusion} and Eqn.~\ref{eqn:windhorst}, the brighter sources that generate these artefacts also tend to be large.

\begin{figure}
    \centering
    \includegraphics[width=0.45\textwidth]{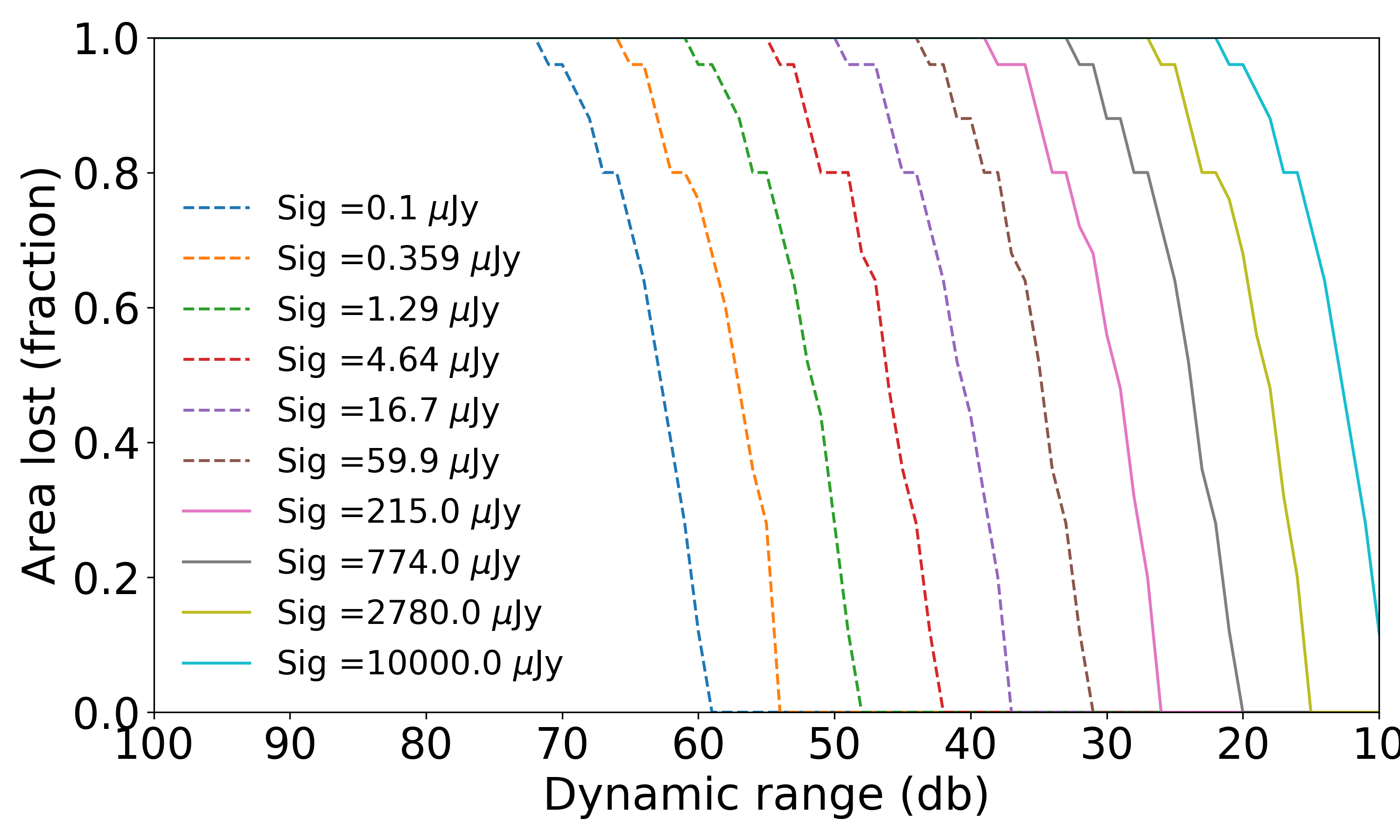}
    \caption{Fraction of the sky that will be inaccessible as a function of dynamic range for size different flux limits (\textit{Sig}), using the distribution of radio galaxy fluxes taken from the  T-RECS simulated catalogue \citep{2019MNRAS.482....2B}. As the dynamic range decreases, the number of bright sources that needs to be masked increases, until the masks dominate the sky area. It is clear that to achieve an all-sky survey of radio continuum galaxies with a a sub $\mu$Jy flux density, a dynamic range of at least 70dB ($10^7$ in flux ratio) would be required. }
    \label{fig:fksy_dbcut}
\end{figure}

Figure \ref{fig:fksy_dbcut} shows the relationship between sky coverage and dynamic range for a number of different flux limits. Here we assume that we must remove about 1 sq. deg around each bright source above the flux maximum $S_{\rm max}$, as the CLEAN algorithm has not been significantly improved on for this action. 

\section{Next generation continuum surveys}
\label{sec:design_options}

\begin{table*}
\centering
\caption{\label{tab:future_surveys}Survey parameters of the current  and future planned radio continuum surveys, based on exiting data sets and future plans. The appropriate references are found in the text. Here we define the optimal number of redshift bins based on those that give less than 10\% shot noise contribution to the error budget. }
\begin{tabular}{ l l c c c c c}
\hline\hline
 Survey & Observatory & Frequency (MHz) & Flux limit & $f_{\rm sky}$ & Optimal $n_{\rm bins}$  \\ 
 \hline
 NVSS & VLA & 1350-1450 & 2.5mJy  &  0.65 & -- \\
 LoTSS & LoFAR & 120-168 & 500 $\mu$Jy & 0.65 & -- \\  
 EMU & ASKAP &  800-1400  & 50 $\mu$Jy  & 0.65 & 1  \\
 MeerKLASS & MeerKAT & 900-1670 & 25  $\mu$Jy & 0.1 & 1 \\
 SKA-MID Phase I & SKA-MID B2   & 950-1750 & 8 $\mu$Jy & 0.12 & 2 \\
 SKA-Survey Phase II & ?  & 800-1400 & 1 $\mu$Jy & 0.65  &  4\\
 \hline
\end{tabular}

\end{table*}

Table \ref{tab:future_surveys} summarises the current and future radio continuum clustering surveys. These are: the NRAO VLA Sky Survey (NVSS, \cite{NVSS}), the  LOFAR Two-metre Sky Survey (LoTSS, \cite{2017A&A...598A.104S,2019A&A...622A...1S}), the Evolutionary Map of the Universe survey (EMU, \cite{2011PASA...28..215N}), the MeerKAT Large Area Synoptic Survey (MeerKLASS, \cite{MeerKLASS}), and the Square Kilometer Array Phase I medium cosmology survey (SKA, \cite{SKA2015Jarvis,2020PASA...37....7S}). Since SKA Phase I does not contain a survey telescope, conducting a large-area continuum survey with SKA-MID will be more time-consuming, due to the smaller field-of-view. Therefore the SKA Cosmology Red book \citep{2020PASA...37....7S} baseline continuum survey is designed to cover only 5000 square degrees, and the sample will be also used for radio weak lensing shear analysis.

A true all-sky radio continuum survey with SKA, which can achieve a significant improvement on the EMU depth, may therefore need to wait until SKA Phase II. In table \ref{tab:future_surveys} we list such a successor survey, with a proposed flux limit of 1 $\mu$Jy. This future survey would require an angular resolution of at least 1 arcsec, an integration time of at least two hours per field of view (assuming a 50K system temperature), and a dynamic range of at least 60dB. This would allow us to achieve sample-variance limited measurements of the clustering power spectrum in 8 redshift bins, assuming a single population (from Fig.~\ref{fig:flux_limit_snr}), and reasonably small shot noise contribution for a multi-tracer analysis for a single bin (from right panel in Fig.~\ref{fig:flux_limit_snr}).

Beyond this, it would be possible to extend down further to a $100 n$Jy limit, which would still only need an angular resolution of 1 arcsec (to avoid confusion), a dynamic range of 70dB, and a significant improvement in system temperature. But, given the results from from Fig.~\ref{fig:natural_confusion}, any fainter than this, and the sources would be confused, overlapping due to their natural size. This natural confusion limit, which is generated by a population of faint but large `monster' AGN (if they are present), would not be resolvable by any improvement in angular resolution by increasing the length and number of long baselines. The next generation of radio continuum surveys, particularly very deep radio continuum observations from the SKA, would reveal or falsify such a population.

\section{Conclusions}

In this paper we have studied the limitations of current and future radio continuum surveys, from the perspective of measuring the angular clustering statistics of radio galaxies to determine the large-scale structure of the Universe. We considered how sources of noise (both sky, and instrumental) can limit the possible flux limit, and so the number of available sky sources, and the region of sky available to survey. This limit on the number of sources then has consequences for how precisely the angular clustering statistics can be measured, especially in the cases where the sample is sub-divided by redshift and species. We summarise our conclusions as follows:

\begin{itemize}
    \item The shot noise term in the clustering error measurement can be reduced by going deeper, which adds more galaxies to the sample. Surveys such as EMU \& MeerKLASS should reach the $\sim 100\mu$Jy flux limit needed for the shot noise to be sub-dominant ($<10\%$ of the total) in a single redshift bin.
    \item For multiple redshift bins, the sample is sub-divided, and so for a given flux the number density will be lower in each individual bin. The survey will need to go even deeper in flux to reduce the shot noise, and to achieve a $<10\%$ shot noise contribution in at least six redshift bins, a flux limit of roughly $500$nJy, corresponding to a $50$nJy rms, would be needed. 
    \item For a survey such that only reaches a $\sim 100\mu$Jy limit, the multi-tracer approach will be not very effective, as the shot noise for individual populations is too large (as was previously shown in \citet{2019JCAP...02..030B}).
    \item Considering the less numerous AGN population as the limiting factor for a multi-tracer analysis, the shot noise is never smaller than $10\%$ of the sample variance, even for a single redshift bin. There are simply not enough AGN available in the sky (given current modelling of the population) to achieve the required accuracy.
    \item To reach the required flux limits for a usable cosmological sample, instrument sensitivity (system temperature) is now less important than angular resolution, needed to minimise the confusion noise, and dynamic range, necessary to maximise the sky area. 
    \item The requirements for a future large-scale survey with a flux limit of around 100$n$Jy (needed for shot noise $<10\%$ in six redshift bins), the telescope needs to operate at 70dB with a angular resolution of 1''. So while 50K detectors, such as the Phased Array Feeds of ASKAP, would be sensitive enough, more detectors with longer baselines and a greater data processing capacity would be needed for such a future survey. Such would be the requirement for any SKA survey instrument for Phase II.
    \item The T-RECS \citep{2019MNRAS.482....2B} simulation predicts a population of 'monster' AGN with median size greater than 1 arcsec  hiding at  depth of 100$n$Jy. It will not be possible to survey to much greater depths than this over a wider area because of these, as the natural confusion of overlapping sources will make surveys deeper than that impossible.  If these exist, they provide a hard limit beyond which it would not be possible to increase the number density by longer observations.
\end{itemize}

The next generation of large-area radio surveys, such as an SKA-Survey instrument as part of Phase II, would have the capabilities needed to measure the angular correlations of galaxies in multiple redshift bins with a small shot noise contribution to the error. Such measurements will provide a very  useful data set to learn more about cosmology and fundamental physics.  But for the  generation of radio telescopes that come after SKA, and plan to measure the clustering of even fainter radio continuum galaxies over a large area of the sky,  there will be challenges, and both in terms of instrumental limitations and the nature of the radio galaxy population. Such a survey may not produce any significant improvement in what will by then have already been measured.

\section*{Data Availability Statement}

The data underlying this article were accessed from the Tiered Radio Extragalactic Continuum Simulation (T-RECS),  accessed from \url{http://cdsarc.u-strasbg.fr/viz-bin/qcat?VII/282}. The derived data generated in this research will be shared on reasonable request to the corresponding author.

\section*{Acknowledgements}

We would like to thank Ian Harrison and Jeffrey Hodgson for helpful discussions when preparing this paper. We would like to thank the EMU cosmology working group for their continued support, including Ray Norris, Glen Rees, Song Chen and Faisal Rahman. We also thank the referee for helpful comments that contributed to the paper. JA has received funding from the European Union’s Horizon 2020 research and innovation programme under grant agreement No. 776247 EWC. DP is supported by the project \begin{CJK}{UTF8}{mj}우주거대구조를 이용한 암흑우주 연구\end{CJK} (``Understanding Dark Universe Using Large Scale Structure of the Universe''), funded by the Ministry of Science.

This research made use of Astropy,\footnote{http://www.astropy.org} a community-developed core Python package for Astronomy \citep{astropy:2013, astropy:2018}, the HEALPix and Healpy package \citep{2005ApJ...622..759G,Zonca2019}, the  Numpy package \citep{book}, the Scipy package \citep{2019arXiv190710121V} and Matplotlib package \citep{Hunter:2007}.

\bibliographystyle{mnras}
\bibliography{refs}

\appendix


%


\section{Bias of an ensemble}
\label{appendix:bias_total}

The bias is an ansatz that relates the number density of galaxies or other tracers to that of matter
\begin{equation}
\label{eqn:bias_def}
    \delta_g(x) = \frac{n_g(x)}{\bar{n}_g} -1 = b\delta_m(x)\,,
\end{equation}
where $\delta_g(x)$ is the normalised galaxy number density fluctuation at some position in space $x$, $n_g(x)$ is the galaxy number density at that position, $\delta_m(x)$ is the matter density fluctuation at that position, $\bar{n}_g$ is the homogeneous number density, and $b$ is the bias.

If we have multiple galaxy species, each with its own bias value (i.e. galaxies that have formed differently and trace the underlying matter density in a different way), how can we estimate the total bias of the ensemble? Firstly we write down what the fluctuation will be in terms of the combined ensemble of tracers, and some total bias,
\begin{equation}
    \frac{n_g^A(x) + n_g^B(x) + \ldots + n_g^X(x)}{\bar{n}_g^A + \bar{n}_g^B + \ldots + \bar{n}_g^X} -1 = b_{\mathrm{total}} \delta_m(x)\,,
\end{equation}
Now we assume a form of the total bias $b_{\mathrm{total}}$, such that the total is the sum of the individual biases $\{b^A,~b^B,~\ldots, b^X\}$ in the form
\begin{equation}
    b_{\mathrm{total}} = \frac{b^A\bar{n}_g^A + b^B\bar{n}_g^B + \ldots + b^X\bar{n}_g^X}{\bar{n}_g^A + \bar{n}_g^B + \ldots + \bar{n}_g^X} \,.
\end{equation}
It can therefore easily be shown that this is the only form of the total bias that satisfies the definition of the linear bias given in Eqn.~\ref{eqn:bias_def} for any and all of the tracers when considered individually.






\bsp	
\label{lastpage}
\end{document}